\documentclass[conference]{IEEEtran}
\ifCLASSOPTIONcompsoc
  \usepackage[nocompress]{cite}
\else
  \usepackage{cite}
\fi
\ifCLASSINFOpdf
\else
\fi
\usepackage{balance}  
\usepackage{graphics} 
\usepackage{txfonts}
\usepackage{times}    
\usepackage[pdftex]{hyperref}
\usepackage{color}
\usepackage{textcomp}
\usepackage{booktabs}
\usepackage{ccicons}

\usepackage{cite}
\usepackage{url}
\usepackage{fancybox}
\usepackage{multirow}
\usepackage{flushend}
\usepackage{booktabs}
\usepackage{tabularx}
\usepackage{comment}
\usepackage{array}
\usepackage[flushleft]{threeparttable}
\usepackage{mdframed}
\graphicspath{{}{images/}{dia/}}
\DeclareGraphicsExtensions{.pdf,.png,.svg}

\usepackage{listings}
\usepackage{courier}
\usepackage{hyperref}

\usepackage{xpatch}

\xpatchcmd{\refstepcounter}{%
  \stepcounter{#1}%
}{%
  \stepcounter{#1}%
  \xdef\scr{\number\value{#1}}%
}{\typeout{success}}{\typeout{failure}}

\newcounter{o}
\usepackage{tikz}
\definecolor{1c1}{RGB}{188,162,6}
\definecolor{1c2}{RGB}{137,129,80}
\definecolor{1c3}{RGB}{239,167,31}
\definecolor{1c4}{RGB}{88,194,241}
\definecolor{1c5}{RGB}{6,180,188}

\tikzset{mynode/.style={draw=white,solid,circle,fill=green,inner sep=1pt, thick,
text=black}}
\tikzset{arrow line/.style={dashed, line width= 2.5pt, color=#1}}
\usepackage{varwidth}
\def\bf{\textbf}

\def\fig {Figure~}

\def\tbl {Table~}

\def\sec {Section~}

\def\it{\textit}

\def\tt{\mct}
\newcommand{\ib}[1]{{\textbf {\textit { #1}}}}

\newcommand{\mct}[1]{{\footnotesize {\texttt {#1}}}}

\usepackage{paralist}

\newcommand{\nd}{\vspace{1mm}\noindent}
\usepackage[small,bf]{caption}
\usepackage{tikz}

\lstdefinestyle{inlinecode}{basicstyle={\ttfamily\scriptsize\bfseries}}

\newcommand{\urls}[1]{{\scriptsize\url{#1}}}
\usepackage{tcolorbox}
\newcommand{\emt}[1]{\emph{``#1''}}

\usepackage{paralist}
\usepackage[outercaption]{sidecap}
\usepackage [autostyle, english = american]{csquotes}
\MakeOuterQuote{"}
\newcounter{scn}
\usepackage[shortlabels]{enumitem}
\usepackage{tikz}
\usepackage{styles/pgf-pie}
\usepackage{paralist}
\usepackage{soul}

\usetikzlibrary{positioning,shadows}
\usepackage{balance}
\newcommand{\dq}[1]{\href{https://stackoverflow.com/questions/#1/}{$Q_{#1}$}}
\newcommand{\da}[1]{\href{https://stackoverflow.com/answers/#1/}{$A_{#1}$}}

\newif\ifpienumberinlegend
\pgfkeys{/number in legend/.code=
    \expandafter\let\expandafter\ifpienumberinlegend
    \csname if#1\endcsname
    \ifpienumberinlegend

    \def\beforenumber##1\afternumber{}%
    \fi,
    /number in legend/.default=true
}
\definecolor{1c1}{RGB}{188,162,6}
\definecolor{1c2}{RGB}{137,129,80}
\definecolor{1c3}{RGB}{239,167,31}
\definecolor{1c4}{RGB}{88,194,241}
\definecolor{1c5}{RGB}{6,180,188}

\tikzset{mynode/.style={draw=white,solid,circle,fill=green,inner sep=1pt, thick,
text=black}}
\tikzset{arrow line/.style={dashed, line width= 2.5pt, color=#1}}

\usepackage{bchart}
\begin{document}
\title{Security and Machine Learning Adoption in IoT: A Preliminary Study of IoT Developer Discussions}
\author{Gias Uddin, DISA Lab, University of Calgary, Canada}

\IEEEtitleabstractindextext{%
\begin{abstract}
Internet of Things (IoT) is defined as the connection between places and physical objects (i.e., things) over the internet/network 
via smart computing devices. IoT is a rapidly emerging paradigm that now encompasses almost every aspect of our modern life. 
As such, it is crucial to ensure IoT devices follow strict security requirements. At the same time, the prevalence of IoT devices offers developers 
a chance to design and develop Machine Learning (ML)-based intelligent software systems using their IoT devices. 
However, given the diversity of IoT devices, IoT developers may find it challenging to introduce appropriate 
security and ML techniques into their devices. Traditionally, we learn about the IoT ecosystem/problems by conducting surveys of IoT developers/practitioners. 
Another way to learn is by analyzing IoT developer discussions in popular online developer forums like Stack Overflow (SO). 
However, we are aware of no such studies that focused on IoT developers' security and ML-related discussions in SO. 
This paper offers the results of preliminary study of IoT developer discussions in SO. First, we collect 
around 53K IoT posts (questions + accepted answers) from SO. Second, we tokenize each post into sentences. Third, we 
automatically identify sentences containing security and ML-related discussions. 
We find around 12\% of sentences contain security discussions, while around 0.12\% sentences contain ML-related discussions. There is no overlap 
between security and ML-related discussions, i.e., IoT developers discussing security requirements did not discuss ML requirements and vice versa. 
We find that IoT developers discussing security issues frequently inquired about how 
the shared data can be stored, shared, and transferred securely across IoT devices and users. We also find that IoT developers are 
interested to adopt deep neural network-based 
ML models into their IoT devices, but they find it challenging to accommodate those into their 
resource-constrained IoT devices. Our findings 
offer implications for IoT vendors and researchers to develop and design novel techniques for improved security and ML adoption into IoT devices.              
\end{abstract}

\begin{IEEEkeywords}
IoT, Security, Machine Learning, Developer Discussions.
\end{IEEEkeywords}}

%


\maketitle

\IEEEdisplaynontitleabstractindextext
\IEEEpeerreviewmaketitle

\section{Introduction}
Internet of Things (IoT) is a rapidly emerging paradigm that is defined as the connection between places and physical objects (i.e., things) over the 
Internet/network via smart computing 
devices~\cite{Atzori-SurveyIoT-ComputerNetwork2010,Gubbi-IoTVisionDirection-FGCS2013}. This technological revolution now encompasses almost 
every aspect of our modern life and is not showing any signs of slowing down to evolve into new domains (e.g., smart cars, smart home, etc.)~\cite{Fuqaha-IoTSurveyTechnologiesApplications-IEEECST2015,Pretz-TheNextEvolutionInternet-IEEEMagazie2013}.
Indeed, between 2013 and 2020, the number of smart connected devices has increased by more than 1000\%, from 5 billion to more than 50 billion~\cite{Chase-EvolutionIoT-TexasInstruments2013}. 
As such, interests in IoT technologies is pervasive among developers to develop smart connected software ecosystems~\cite{Weyrich-RefArchitectureIoT-IEEESoftware2016}.

The access to diverse and large sensor data generated by the IoT devices have created opportunities to 
adopt Machine Learning (ML) into IoT-based solutions~\cite{Marjani-IoTDataAnalytics-IEEEAccess2017}. At the same time, the pervasiveness of the IoT devices 
in our everyday life has necessitated the developers to adopt security techniques and tools into their IoT devices. 
~\cite{Chi-SmartHomeCrossAppInference-DSN2020,Ding-IoTSafetyPhysicalInterfaction-CCS2018,Edwards-IoTHajimeWorm-Rapidly2016,Wang-FearLoggingIoT-NDSS2018,Sekar-HandlingTrillionIoTSecurity-HotNets2015}. 
Security concerns for IoT devices can be multifaceted like the 
implementation/adoption of security protocols (e.g., zigbee chain reaction~\cite{Ronen-IoTNuclearZigbeeChainReaction-SP2017}) and 
roles (e.g., authentication~\cite{Gong-IoTPIANO-ICDCS2017}), the communication over the network like (e.g., cross-device inference~\cite{Chi-SmartHomeCrossAppInference-DSN2020}), 
as well as the underlying hardware~\cite{Ho-IoTSmartLocks-ASIACCS2016}. Indeed, adoption of security and ML techniques into 
low-powered but omnipresent IoT devices requires efforts of unprecedented nature 
unlike anything we have seen before~\cite{Sekar-HandlingTrillionIoTSecurity-HotNets2015}. As such, it is important to understand 
the challenges IoT developers face during their adoption of security and ML practices, so that we can 
design effective techniques to address the challenges. 
 
With interests in IoT growing, we observe discussions of IoT developers in online forums like Stack Overflow (SO). 
To date, there are
around 120 million posts and 12 million registered users on Stack
Overflow~\cite{website:stackoverflow}. 
Previously, several research has been conducted to
analyze SO posts, e.g., to analyze discussions on big
data~\cite{Bagherzadeh2019}, concurrency~\cite{Ahmed-ConcurrencyTopic-ESEM2018},
programming issues~\cite{Barua-StackoverflowTopics-ESE2012}, blockchain
development~\cite{wan2019discussed}, microservices~\cite{bandeira2019we}, and
security~\cite{yang2016security}. However, we are aware of no
research that analyzed IoT security and ML-related discussions on SO, although such insight can complement existing IoT literature that so far has mainly used 
surveys~\cite{Atzori-SurveyIoT-ComputerNetwork2010,Gubbi-IoTVisionDirection-FGCS2013,Fuqaha-IoTSurveyTechnologiesApplications-IEEECST2015}.  

In this paper, we present a preliminary empirical study that we conducted to understand security and ML adoption in IoT devices based on the discussions of IoT 
developers in SO. First, we collect 53K IoT posts from SO based on 78 IoT tags (\sec\ref{sec:data}). 
Second, we tokenize the posts into sentences, 
which resulted in around 672K sentences. 
Third, we automatically label each sentence with two labels: \begin{inparaenum} 
\item `HasSecurity' is 1 if the sentence contains security discussions and 0 otherwise, and 
\item `HasML' is 1 if the sentence contains ML-related discussions and 0 otherwise.  
\end{inparaenum} We use the labels to answer two research questions (\sec\ref{sec:results}):

\nd\bf{RQ1. How do developers discuss security issues while using IoT techniques and tools?} 
This question focuses on understanding security issues from IoT developer discussions in SO. We find that around 12\% of all sentences 
contain discussions about security, i.e., IoT developers frequently are concerned about security requirements and problems in their IoT devices. 
Their security concerns are multifaceted like involving the secure access/transmission of data across IoT devices and users, and the diverse errors 
and incompatibilities they experience while enforcing security protocols across the diverse IoT devices.

\nd\bf{RQ2. How do IoT developers discuss machine learning issues and is there any overlap with security issues?}
This question aims to understand how IoT developers are adopting ML-specific techniques/services into their IoT devices. We observed 0.12\% of sentences 
containing ML-related discussions, i.e., the ML-discussions are less prevalent than security discussions. There is no overlap between the security and ML discussions, 
i.e., IoT developers in SO working on security may not be working on ML adoption at the same time.  We find that IoT developers are interested to adopt deep neural network-based 
ML models into their IoT devices, but they find it challenging to accommodate those into their resource-constrained IoT devices. 

Our findings 
offer implications for IoT vendors and researchers to develop and design novel techniques for improved security and ML adoption into IoT devices.   

\nd\bf{Replication Package}: \urls{https://github.com/disa-lab/SERP4IoT2021}

\section{Data Collection}\label{sec:data}
We follow three steps to collect SO posts related to IoT discussions: \begin{inparaenum}[(1)]
\item Download SO dataset,
\item Identify IoT tagset in the dataset, and
\item Download IoT posts from the dataset based on the IoT tagset.
\end{inparaenum} We describe the steps below.

\nd\bf{\ul{Step 1. Download SO Dataset.}} 
We download the SO data dump~\cite{website:stackoverflow-datadump} 
of September 2019 and use the Posts table in the dump for our analysis. 
A post can be a question or an answer in the Posts table. An answer to a question is flagged as accepted, if the user 
who has asked the question marked the answer as accepted. A question can have between 1 and 5 tags.
The SO  dataset includes all posts for 11 years between 2008 
and September 2019.  In total, it has 46,767,375 questions and answers. 
Out of those around 40\% are questions and 60\% are answers. 
Out of the answers, around 34\% are accepted. 

\nd\bf{\ul{Step 2. Develop IoT Tag Set.}} 
We use tags in SO to identify IoT posts based on an algorithm originally proposed by 
Yang et al.~\cite{YangLo-SecurityDiscussionsSO-JCST2016}. The approach starts with a list of initial tags as seed data. 
Then the approach uses two metrics to automatically expand the list of tags. We discuss each step below.

First, we identify a list of initial IoT tags in SO that 
frequently co-occurred with the `iot' tag. This 
resulted in a list of 20 relevant tags based on SO search engine like `raspberry-pi', `arduino', `windows-10-iot-core', `python', etc. 
From this list, 
we removed potentially irrelevant tags like `python'. We found that the rest of the tags can 
be clustered into three broad types: \begin{inparaenum}
\item iot or any tag with `iot' keyword, e.g., `azure-iot-hub', 
\item arduino, and 
\item raspberry-pi.
\end{inparaenum} Besides the iot tags, arduino and raspberry-pi are the two most popular open source 
platforms to develop IoT based applications (with dedicated SDKs). Both platforms 
have undergone rapid evolution through multiple versions, such as raspberry-pi, raspberry-pi2, etc. We note the 
list of initial tags as $\mathcal{T}_{init}$.         

Second, we check the entire list of SO tags that could be relevant to the three tags in $\mathcal{T}_{init}$. 
Suppose, the entire SO data dump is denoted as $\mathcal{D}$ and the list of all tags in the SO data dump is $\mathcal{A}$. 
We extract a list of all questions $\mathcal{P}$ from our dataset that are labeled as at least one of those tags in $\mathcal{T}_{init}$. 
Not all the tags in $\mathcal{A}$ may correspond to IoT (e.g., python). 
Therefore, following previous research~\cite{YangLo-SecurityDiscussionsSO-JCST2016,Bagherzadeh-BigdataTopic-FSE2019} 
we systematically filter out \it{irrelevant} tags and 
finalize a list of all potential IoT tags $\mathcal{T}$ 
from $\mathcal{A}$ as follows. For each tag $t$ in $\mathcal{A}$, we compute its significance and relevance 
with regards to $\mathcal{P}$ and $\mathcal{D}$.  
\begin{equation}
\textrm{Significance}~\mu(t) = \frac{\# \textrm{of Questions with tag $t$ in $\mathcal{P}$}}{\# \textrm{of Questions with tag $t$ in $\mathcal{D}$}}
\end{equation}
\begin{equation}
\textrm{Relevance}~\nu(t) = \frac{\# \textrm{of Questions with tag $t$ in $\mathcal{P}$}}{\# \textrm{of Questions in $\mathcal{P}$}}
\end{equation}
A tag $t$ is significantly relevant to IoT discussions if $\mu(t)$ and $\nu(t)$ are higher than a specific threshold. 
Our 49 experiments with a 
broad range of threshold values of $\mu = \{0.05, 0.1, 0.15, 0.2, 0.25, 0.3, 0.35\}$ and $\nu = \{0.001, 0.005,
0.01, 0.015, 0.02, 0.025, 0.03\}$ show that $\mu = 0.3$ and $\nu = 0.001$ allow for a significantly 
relevant set of 78 IoT tags. The threshold values 
are consistent with 
previous work~\cite{YangLo-SecurityDiscussionsSO-JCST2016,Ahmed-ConcurrencyTopic-ESEM2018,Bagherzadeh-BigdataTopic-FSE2019}. 
The 78 IoT tags 
cover a wide range technologies and services supporting the emerging IoT ecosystems. 
Our replication package lists 78 IoT tags.

\nd\bf{\ul{Step 3. Download IoT Posts.}} Our final dataset consists of all posts tagged as at least one 
of the candidate 78 tags. We found a total of 81,651 posts (questions and answers) in our SO dump, 
out of which around 48\% are questions (i.e., 39,305) and 52\% (42,346) are answers. Following previous
research~\cite{Bagherzadeh-BigdataTopic-FSE2019,Barua-StackoverflowTopics-ESE2012,Rosen-MobileDiscussionSO-EMSE2016},
we only consider the questions and accepted answers to the questions.
Our final dataset consists 53,173 posts (39,305
questions, 13,868 accepted answers).

\section{Empirical Study} \label{sec:results}
\nd We answer two research questions using the 53K IoT posts:
\begin{enumerate}[leftmargin=30pt, label=\bf{RQ\arabic{*}.}]
  \item How do developers discuss security issues while using IoT techniques and tools?
  \item How do IoT developers discuss machine learning issues and is there any overlap with security issues?
\end{enumerate} 
The first question (RQ1) focuses on understanding security issues in IoT as developers discuss those in SO. 
The second question (RQ2) aims to understand how IoT developers are using/adopting ML 
into IoT devices, which are often resource constrained like low computing storage/processing power, etc. While the adoption of 
security and ML into IoT devices is gaining attention, it is not known whether both requirements are considered simultaneously and whether 
IoT developers discuss those two requirements together in SO. 

\subsection{IoT Security Issues in Developer Discussions (RQ1)}
\subsubsection{Approach} Not all discussions in an SO post may be related to security. We therefore detect sentences in SO that contain security discussions as follows. 
First, we produce a benchmark dataset of 5,297 sentences from SO, each labeled as 1 (contains security discussion) or 0 (otherwise). Out of the 5,297 sentences, 
4,297 sentences are taken from a benchmark of API reviews previously developed by Uddin and Khomh to develop the Opiner tool~\cite{Uddin-OpinionValue-TSE2019}. Opiner is 
an online portal to mine and summarize reviews from SO about diverse API aspects (e.g., performance, security, etc.). Each sentence 
in the Opiner benchmark is labeled as one or more API aspect. The Opiner dataset does not include any sentence 
related to IoT discussions. We, therefore, randomly sampled 1,000 sentences from our IoT dataset of 53K posts. We do this by tokenizing each post into sentences and then 
sampling 1,000 sentences from the sentences. Second, we trained a suite of shallow and deep learning models on the benchmark of 5,297 sentences. Third, 
we evaluate the performance of the models to correct detect security-related sentences in another validation dataset of 984 sentences. The validation dataset 
is sampled from our 53K IoT posts. The validation and training (i.e., benchmark) datasets are mutually exclusive. The best performing classifier was RoBERTa, 
an advanced pre-trained language-based models, which shows an F1-score of 0.91 (Precision 0.97, Recall 0.87). The high accuracy 
of the model denotes that the model can be reliably used to automatically collect IoT developers' security discussions. Fourth, we applied the trained RoBERTa~\cite{Liu-Roberta-Arxiv2019} model 
on our entire 53K IoT posts to automatically label as sentence as 1 (i.e., contains security discussion) or 0. The details of the benchmark creation process and 
the models can be found in our technical report~\cite{Nibir-DeepIoTSecurty-MSR2021}. 
We have also shared  in our replication package the code of RoBERTa and the benchmark that is used to train the model.

\subsubsection{Results} We found total 672,678 sentences in our 53K IoT posts, out of which 30,192 sentences are 
labeled as 1 (i.e., they contain security discussions) by our RoBERTa model.  
In \tbl\ref{tab:secdataCount}, we show the distribution of the security-related sentences in the 53K posts. 
The 30,192 sentences correspond to 4.5\% of our entire dataset. The security-related sentences are found in 5,354 questions (\#Questions in \tbl\ref{tab:secdataCount}), 
which correspond to 13.6\% of all the total 39,305 IoT questions in the dataset. Out of the 13,686 accepted answers in the IoT dataset, 
12.5\% answers contain discussions about security (\#Answers column). The numbers denote that IoT security-related discussions 
are prevalent and frequent in SO. 
\begin{table}[t]
  \centering
  \caption{Distribution of sentences \& posts with security discussions}
    \begin{tabular}{rrrrrr}\toprule
    \#Sentences & \%Total    & \#Questions & \%Total    & \#Answers & \%Total \\
    \midrule
    30,192 & 4.49\% & 5,354  & 13.62\% & 1,734  & 12.50\% \\
    \bottomrule
    \end{tabular}%
  \label{tab:secdataCount}%
\end{table}%

\begin{figure}[t]
\centering
   	\hspace{-5mm}
   	\includegraphics[scale=.6]{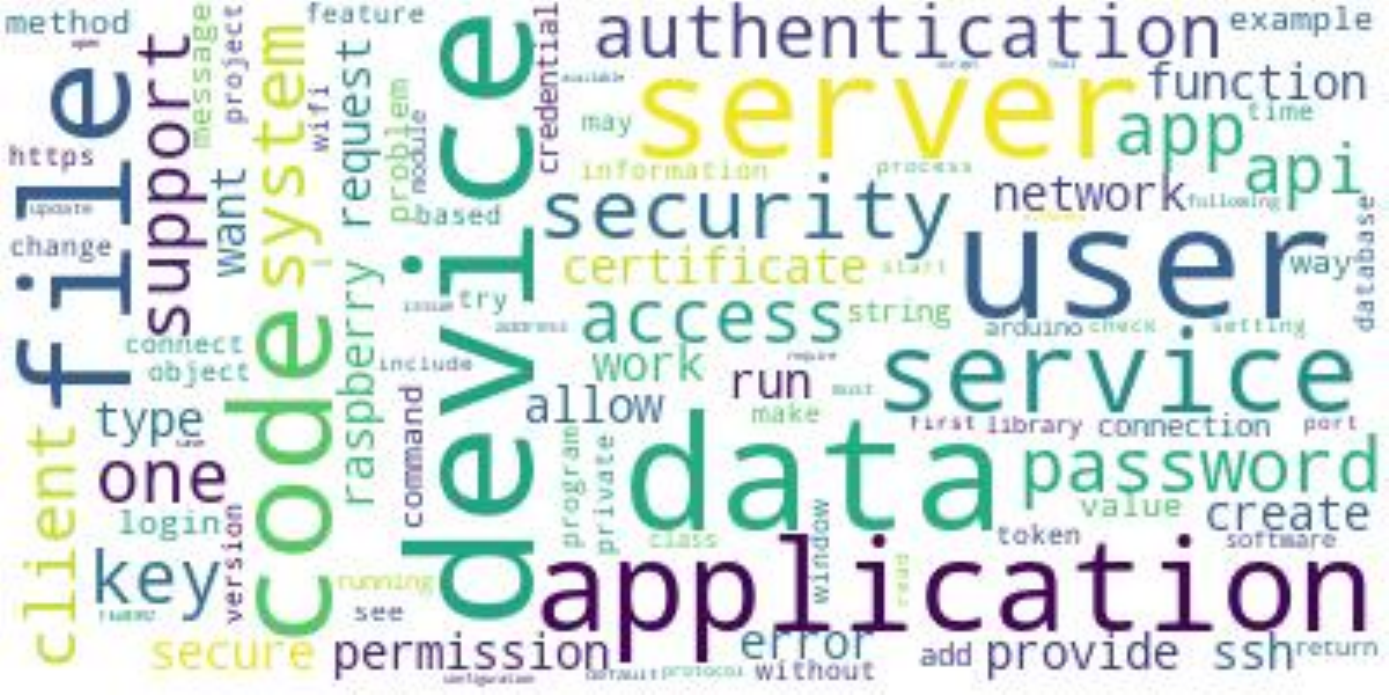}
   	\caption{The frequently occurred keywords in security discussions}
   	 \label{fig:wordcloudSecurity}
\end{figure}

\fig\ref{fig:wordcloudSecurity} shows a wordcloud of top 100 most frequent keywords in the IoT security 
discussions\footnote{Word clouds are created using the \href{https://github.com/amueller/word_cloud}{Python WordCloud API}}. The font size of a word 
is proportionate to its frequency of occurrences relative to all keywords in the  in the 30K security-related sentences. We remove stopwords before 
we generate the wordcloud\footnote{We use stopwords from Python NLTK~\cite{website:nltk}}. The security 
concerns of the IoT developers cover diverse issues as highlighted by the keywords like protection of data, communication between devices and servers, secure execution of 
code, etc. 
The most frequent 15 words in the security discussions are: \begin{inparaenum}
\item data, 
\item device, 
\item user, 
\item security,
\item code, 
\item file,
\item server, 
\item access, 
\item wifi, 
\item password,
\item key, 
\item authentication, 
\item private, 
\item ssh, and 
\item connect. 
\end{inparaenum}. The occurrences of the keywords are not mutually exclusive all the time, i.e, multiple keywords could appear together in one 
sentence. We discuss the keywords with examples below. 

\begin{inparaenum}[(1)]
\item \ib{data (2890)}-related security issues are discussed in various contexts like access control to a data source, encryption, secure 
transformation of data during transmission, etc. For example, developer here asks questions about 
the feasibility of creating random number from microcontroller to apply to
cryptography: \emt{the question is about the feasibility of using
microcontroller-gathered data to generate random numbers that could be applied
soundly to cryptography-- an alternative to relying on a device's entropy.} \dq{10864668}\footnote{$Q_i$ and $A_i$ denote a question or an answer in SO with ID i}

\item\ib{device (2607)} keyword appeared in discussions related to the secure 
connection and/or communication between multiple IoT devices or between a non-IoT and IoT device. For example, this developer wants to 
create a remote-controlled door locker \emt{I'm now using the Arduino + WiFiShield to create a door locker which could control the door lock remotely by a portable device with its browser.} \dq{19891735}

\item\ib{user (2450)} keyword is found users to an IoT device need to
access/user it securely, e.g., \emt{Essentially I want users of the app to be
able to login using an RFID card as opposed to inputting their user name and pin,
but I still have a lot of work to do to get this done.} \dq{12023328}

\item\ib{security (2266)} keyword was used when developers were concerned about security risks of their IoT devices/solutions in general, e.g., \emt{I would like to have some sort of security, in that I don't want anyone else to be able to contact my RPi and operate it.} \dq{14237597}

\item\ib{code (2184)}-related security discussions concerned about the implementation of security practices during coding. 
Developers worried about writing plain-text passwords in source code: \emt{This password is written in the code and I would like to hide the password from malicious readers who have access to the .o files and .hex files.} \dq{10326698}. 
Developers also struggled with migrating legacy code into IoT devices: \emt{but I still don't know how to modify the old code with these OAuth codes to get it working again.} \dq{17262109}

\item\ib{file (2179)} related concerns are discussed to secure the access to a file in an IoT device, e.g., \dq{14678552} \emt{The called python file makes use of the GPIO so sudo is required, right?}\dq{16964983}. 
Developers also discussed errors that occurred during secure file access, e.g., 
\emt{When I try to link my sql file (lahman2012.sql) to the database I get an error even though I use the password above (password)}.

\item\ib{server (2105)} related security discussions situated around the the setup of client-server architecture involving multiple IoT devices. For example, 
the developer wants to open a client/server socket over an Arduino serial connection: \emt{I would like to open a client or server socket on the processing program so i can talk to it via WiFi and control my Arduino over serial connection.}\dq{10907872}
Developers also discussed about the potential threats of exposing private keys across similar IoT devices in a network, because if the server is compromised all the IoT devices will be 
compromised:
\emt{But I am wondering if it is a threat to issue the same private key to many devices, that if one device's key is stolen, the whole service on the server side is compromised.} \dq{16329069 }

\item\ib{access (2088)}-related issues can arise due to the secure use of a service/resource. Incompatibilities can arise when 
the authentication is designed for non-IoT environment (e.g., Windows): \emt{We have a web page that we want to access from a raspberry pi, however the web page in question is protected with windows authentication.}\dq{15165212} 
Access problems also occurred when developers attempted to executed shell/python script over an IoT device (e.g., RPI) through the secure connection (e.g., SSH): 
\emt{I have a Raspberry pi that I have been accessing through SSH, but now I need to run a python program on it that has a GUI.}\dq{15712403} 
 
\item\ib{wifi (1929)} issues are dominant when developers attempted secure protocols over wifi 
connection or to dynamically configure the wifi network configuration in an IoT device: 
\emt{My problem is that i want to create a script on my rpi to automatically change wifi networks and change eth0 between static and dhcp.}\dq{12646278} 
Developers also found IoT software libraries work well in simulation but not in an wifi-enabled secure environment: \emt{The library seems to work but I can't figure out the way to connect to a wifi using a password.}\dq{30361166}

\item\ib{password (1828)} changes can be problematic in IoT devices: \emt{I am using Raspberry Pi (Debian 3.1.9+ armv6l linux) and javaSE 7 and I am trying to change my trustore password by keytool but keytool command is not working So can you please tell me how can i change my trustore password in this environment.}\dq{13579503}
Setting up acces points and the configuration of SSID and passwords in IoT devices are also found as challenging: \emt{I'm trying to setup an Arduino to setup an access point to configure SSID and password for connecting to a Wi-Fi network.}\dq{16746045}

\item\ib{key (1824)} store and change are required to securely load and use apps in IoT devices. This can be challenging like when an IoT device (e.g., smart home system) is configured 
to run third-party software:  \emt{I changed out appkey.c for my appkey as given by spotify but when I run the spshell example and try to login, I get an error}\dq{14555190}  

\item\ib{authentication (1656)} parameters and values can be challenging to pass across IoT devices: \emt{Is there any way to pass this authentication across from the Raspberry pi?}\dq{15165212} 
Challenges related to secure messaging passing are discussed. For example, this developer experienced authentication error and automatic disconnection while 
attempting to establish secure messaging using MQTT: \emt{When I try to connect to nodeMCUwifi using my smart phone (just to test the connection, I have no 
intention of using this system for heavy Internet load, only MQTT messages) I get a message "authentication error occurred" 
even though I  have typed the password correctly, or (in rarer cases)it connects but disconnects immediately.} \dq{34130536}

\item\ib{private (1474)} dev server setup can be challenging in IoT devices due to space limitations: \emt{Now i was curious how much space I have used until now after installing all the packages 
I needed for a private dev server.}\dq{18849592}. Developers worried about 
the access private vs public libraries across IoT devices due to privacy issues: 
\emt{The twist is that I don't want to put this a std Arduino library location, I want it to be a "private" library that's visible to and used by only these two sketches.} \dq{20175003}

\item\ib{ssh (1403)}-based communication can be problematic in IoT devices when IoT ports are not properly configured to support that. 
For example, this developer struggled with the setup of ports to access nginx web server from a raspberry pi device: \emt{I have set my router to port all ports from 1 - 9999 just to see 
if it works but I can't ssh onto my pi using the public ip address only my local one works, and I am also unable to access my nginx web server on my pi.}\dq{14530190} 

\item\ib{connect (1400)} issues are common when IoT developers attempted to connect multiple devices and servers to an IoT devices. 
For example, this developer struggled to securely connect an rpi device to a MySQL database through a pre-configured PHP script:  
\emt{I have a little Raspberry Pi server set up, using Nginx and Raspbian, and 
I downloaded a PHP script to serve as a login handler, except that I can't seem to connect to my MySQL database.}\dq{18938608}
\end{inparaenum}

\begin{tcolorbox}[flushleft upper,boxrule=1pt,arc=0pt,left=0pt,right=0pt,top=0pt,bottom=0pt,colback=white,after=\ignorespacesafterend\par\noindent]
\nd\it{\bf{RQ1. How do developers discuss security issues while using
IoT techniques and tools?}} Around 12\% of questions and answers in our IoT dataset contain at least one sentence with discussions of security. 
The discussed security issues are multifaceted like involving the secure access/transmission of data, the configuration of secure communication among IoT devices, 
the diverse errors and incompatibilities IoT developers face while enforcing security principles across IoT devices, and so on.
\end{tcolorbox} 
\subsection{IoT ML Issues in Developer Discussions (RQ2)}
\subsubsection{Approach} We use a set of keywords to identify ML-related sentences in the total 672,678 sentences of our 53K IoT posts. 
The keywords/phrases are: \begin{inparaenum}
\item machine learning,
\item suervised learning,
\item unsupervised learning,
\item deep learning,
\item reinforcement learning, 
\item neural, and 
\item adverserial.
\end{inparaenum} For phrases with more than one word, we try two combinations like `machine learning' and `machine-learning'. The list of keywords/phrases 
are picked to cover the different variants of machine learning algorithms. First, we make each sentence lowercase. Second, we label a sentence as 1 if it contains at least 
one of the keywords/phrases (i.e., it contains ML discussions). Otherwise, we label the sentence as 0.    
\begin{table}[t]
  \centering
  \caption{Distribution of sentences \& posts with ML discussions}
    \begin{tabular}{rrrrrr}\toprule
    \#Sentences & \%Total    & \#Questions & \%Total    & \#Answers & \%Total \\
    \midrule
    801 & 0.12\% & 57  & 0.15\% & 8  & 0.06\% \\
    \bottomrule
    \end{tabular}%
  \label{tab:mldataCount}%
\end{table}%
\subsubsection{Results} We found total 801 sentences containing ML discussions by IoT developers in SO. \tbl\ref{tab:mldataCount} provides 
summary statistics of the ML-related sentences. The ML discussions accounted for 0.12\% of all 
sentences in our IoT dataset. The sentences are found in 57 questions (i.e., 0.15\% of all questions) and eight accepted answers. 
We do not find any overlap between the ML and security discussions, i.e., no sentence or post in our dataset contained discussions of both security 
and machine learning. While this means that IoT developers in our dataset discussed the two requirements (ML and security) separately, 
it could also be perceived as that the IoT developers in our dataset do not find any overlap between their enforcement of security techniques with their 
adoption of ML algorithms/techniques into their IoT-based solutions. This could also denote that among the IoT developers in our dataset, 
security and ML experts may not be the same developers/personnel in a software 
development team. 
\begin{figure}[t]
\centering
   	\hspace{-6mm}
   	\includegraphics[scale=.48]{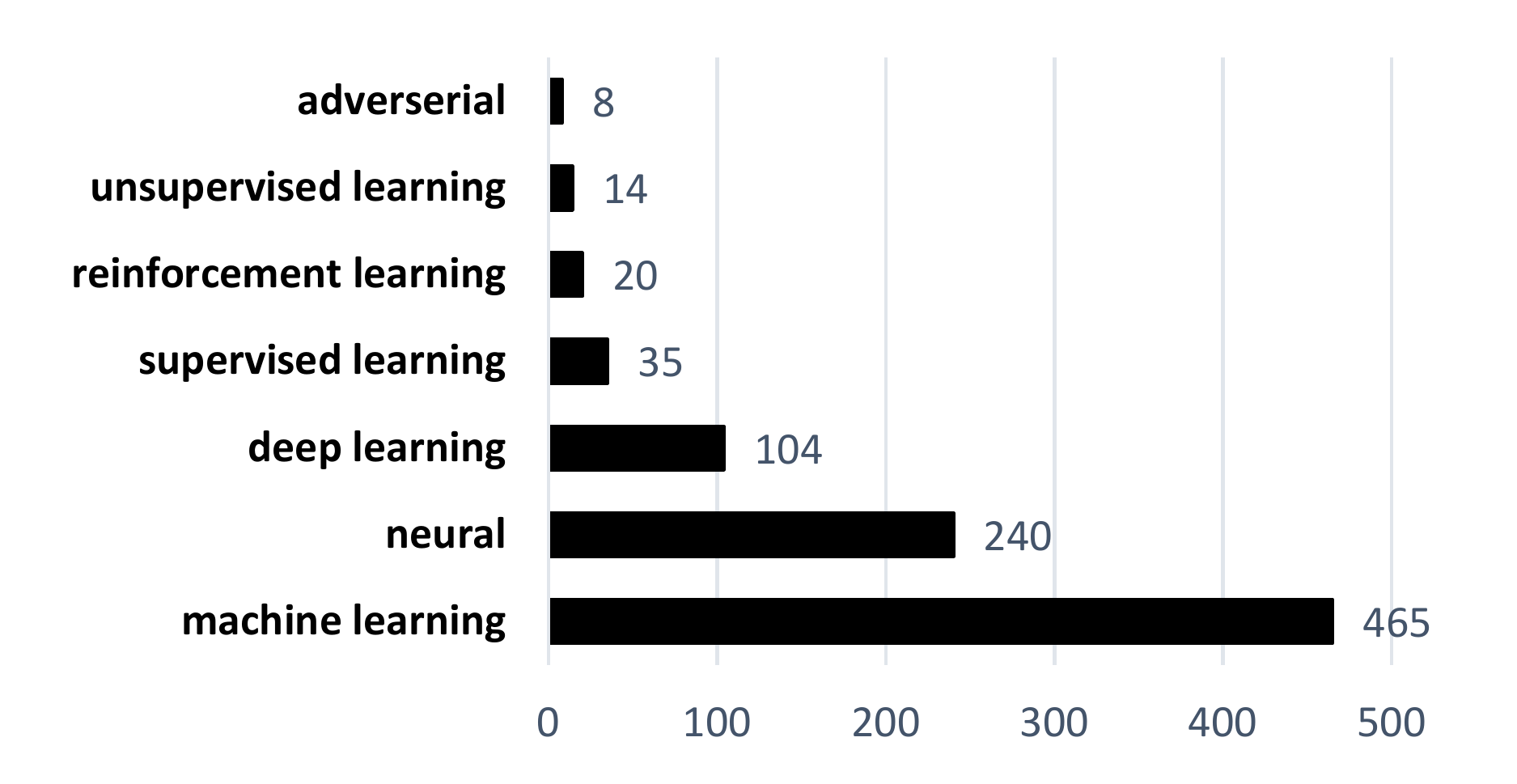}
   	\caption{The ML-keywords used to find sentences discussion ML issues. The number for each keyword denotes the number of sentences}
   	 \label{fig:mlKeywords}
\vspace{-5mm}
\end{figure}

\fig\ref{fig:mlKeywords} shows the distribution of the ML-related sentences by the keywords/phrases we used to find those in our dataset. 
Out of the 7 keywords/phrases, the phrase `machine learning' is found in more than 50\% of the sentences (465), followed by the keyword `neural' (240 sentences). 
Phrases like `supervised learning' and `unsupervised learning' are found less than the phrase `deep learning', but the reason is that discussions about 
traditional shallow learning models (i.e., supervised/unsupervised) are also found under the phrase `machine learning', which developers are more explicit when 
they discuss deep learning algorithms. Therefore, an ML-related sentence can have more than one of the seven search keywords/phrases. 
We discuss below example each keyword/phrases with examples.

\begin{inparaenum}[(1)]
\item \ib{machine learning (465).} While we found majority of ML-related sentences using this phrase, many of these sentences also have one or more of the other keywords/phrases. 
In general, IoT developers discussed machine learning to train/fit ML models into their IoT devices. For example, this developer inquired about the feasibility of using Python 
scikit-learn library to train ML models into a Raspberry Pi2 (RPI2) device: \emt{I am trying to utilize the Python library Scikit-Learn on my Raspberry Pi 2 for machine learning.} \dq{40966243} 
Image/video stream analysis is area where IoT devices (e.g., a smart camera) could be applied in real-world scenarios.  
We observed developers discussing about how they can stream such videos streams from their RPI2 device into a non-IoT device like a Ubuntu ML system:
\emt{ I'm trying to feed my Raspberry Pi MPEG Video stream into a Ubuntu machine learning system called Darknet.} \dq{42824871}
To overcome the low-computing resources in IoT devices that can be problematic for resource-intensive ML models, IoT servers or hubs (e.g., in Azure/Google cloud). 
Developers found it challenging to use such IoT servers for ML tasks, when the documentation is incorrect/incomplete: \emt{ According to https://docs.wso2.com/display/IoTS310/Analyzing+Data I should be able to do some Machine Learning tasks in IoT 
Server but the menu, usually available in WSO2 DAS, is missing, as is the Machine Learner features in "Configure->Features->Installed features" or 
"Configure->Features->Available features".}\dq{46368422}

\item \ib{neural (240).} IoT developers asked diverse questions about the adoption of neural network-based ML models 
into their IoT devices. Such models can be the deep learning models or the ANN (Artificial Neural Network) models found in shallow learning libraries. 
IoT Developers inquired whether they could train a neural network model in a non-IoT device and then deploy the learned model into an IoT device:   
\emt{i would like to train a neural network with tensorflow on my more powerful laptop and then transfer it to the rpi for prediction (as part of a magic mirror).} \dq{40431989} 
The developers also encountered the floating point problem in their code while using neural network model: \emt{I'm trying to calculate a neural network with this code below \ldots The output goes by these value :
-2.0422704 \ldots but when I try to recalculate with matlab, the output goes by (this is the right one):
0.856575444075245 \ldots My question is, is there something wrong with arduino and floating/double calculation which makes the calculation get wrong result?
}\dq{31714225} 

\item \ib{deep learning (104).} IoT developers discussed about training and embedding deep learning model with minimum hardware, so that the learned model could fit into IoT devices 
like Arduino: \emt{Currently I'm trying to replicate this wonderful project \url{https://www.youtube.com/watch?v=u8pwmzUVx48}, 
though with more minimum hardware, in which for the classification i'm using a neural network embedded inside the arduino 
uno which had been trained offline in more stronger PC.} \dq{31800119}
The developers also inquired how the images they are scanning using an IoT device can be fed into a deep learning model: 
\emt{I'd like to feed the scanned images into a deep learning network, such that if I were to hold one or more of my cards in front of a camera, 
it would be able to identify which one(s) I was holding.}\dq{32303587} Questions about the feasibility of using GPUs inside IoT devices are prevalent:
\emt{Can I use gpu for deep learning on raspberry pi 3?}\dq{44381361} 

\item \ib{supervised learning (35), reinforcement learning (20), unsupervised learning (14), adversarial (8).} We found these phrases in mostly answers, where 
IoT developers offered explanation of the types of different machine learning algorithms. Such discussions ranged from high-level 
to API-level (e.g., recommending a specific ML library for a given task). For example, this  
this developer discusses the difference between supervised and unsupervised algorithms: \emt{supervised learning is the machine learning task of inferring a function from labeled training data.}\da{16536040}
This other developer recommends libraries 
to design a supervised learning model:
 \emt{it looks like mpi is simplest to design when it's for supervised learning, or genomics (samecode and large data sets).}\dq{49842838} 
For reinforcement learning, we found similar types of discussions. For example, this developer recommends the OpenAI Gym platform to train reinforcement learning algorithm 
for IoT devices: \emt{openai gym is a platform for reinforcement learning research that aims to provide a general-intelligence benchmark with a wide variety of environments.}\da{40196093}
IoT developers also inquired about the design of adversarial deep neural network in their IoT devices: \emt{deep convolutional generative adversarial networks (dcgan) are ai architectural structures used for convoluted systems, that do not depend on the standard mean-squared error, and so these ai systems learn of their own volition.}\da{49774767}
\end{inparaenum}
\begin{figure}[t]
\centering
   	\hspace{-5mm}
   	\includegraphics[scale=.6]{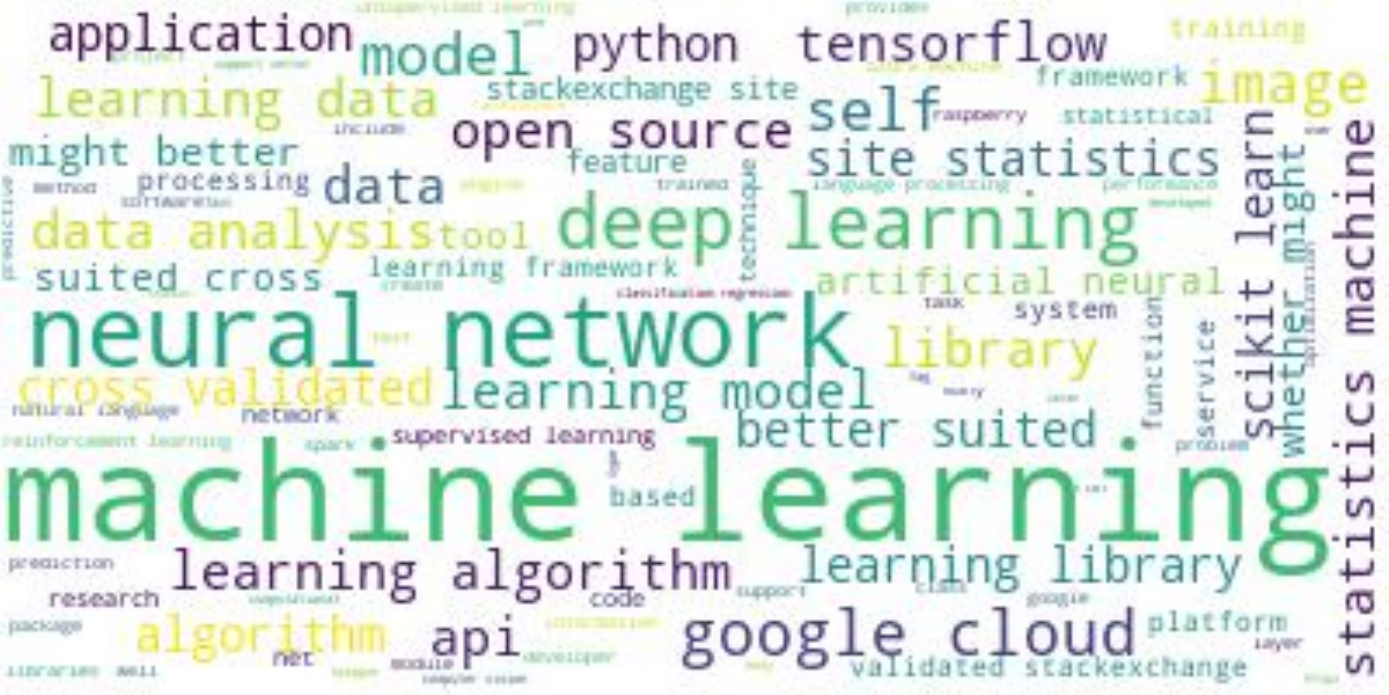}
   	\caption{The frequently occurred keywords in ML discussions}
   	 \label{fig:wordcloudMLUnfiltered}
\end{figure}
In \fig\ref{fig:wordcloudMLUnfiltered}, we show the wordcloud of ML-specific discussions in our dataset. All the keywords/phrases that we used can be seen in the wordcloud, along with 
popular ML libraries like python scikit-learn (for shallow learning) and Google Tensorflow (for deep learning). In addition, we see the phrase `Google Cloud' which is also discussed by IoT 
developers as a cloud-based infrastructure to train and to deploy their ML models into their IoT devices. 
\begin{tcolorbox}[flushleft upper,boxrule=1pt,arc=0pt,left=0pt,right=0pt,top=0pt,bottom=0pt,colback=white,after=\ignorespacesafterend\par\noindent]
\nd\it{\bf{RQ2. How do IoT developers discuss machine learning issues and is there any overlap with security issues?}} Around 0.12\% of the sentences in our dataset contained 
discussions about ML, which is considerably less than the security discussions we observed in our dataset. We also did not find any overlap between the ML and security discussions, i.e., 
IoT developers discussed security and ML-specific requirements in unrelated/different posts. The ML discussions ranged from using traditional shallow learning models to recent resource-intensive 
deep learning and neural network models. For deep learning, developers inquired about the feasibility of using GPUs inside their IoT devices as well as training the models in the cloud 
or in high-powered non-IoT devices and then deploying those trained models into their IoT devices. We also observed many tutorial-like discussions, where developers simply 
offered backgrounds like distinction between ML algorithm types. Therefore, IoT developers in SO are using the Q\&A forum both to learn about their specific programming tasks involving ML models 
as well as to teach each other on the basics of ML domains/algorithms/platforms.   
\end{tcolorbox} 
%
%
%

%
%
%
\section{Discussions} \label{sec:discussion}
\nd\bf{\ul{Implications.}} The development and adoption of the IoT-based solutions by developers 
is facilitated by exponential growth of IoT devices, software, and platforms. Our increasingly inter-connected digital  
world relies on smart devices built using IoT, which means security and ML adoption for IoT devices are paramount 
to leverage IoT-based solutions in wide-range of usage scenarios. The \bf{IoT Vendors} need to support IoT developers 
with proper and usable secure IoT techniques. To determine what is appropriate and what is working, though, the vendors need to 
know the problems faced by IoT developers. 
Such insights can be crucial to the vendors not only to improve their offerings, but also to compare the solutions 
of their competitors. IoT vendors also need to make the IoT devices more usable for ML adoption. In particular, 
the deep neural network requires extensive computing resources, which IoT devices are often not equipped with. Our findings show 
that IoT developers are trying to make their IoT devices smart by adding ML-specific capabilities using deep learning models. However, 
they find it challenging to create a less resource-intensive versions of the ML models that can fit into their IoT devices. Recent 
attempts to create pre-trained language models (e.g., BERT~\cite{Delvin-BERTArch-Arxiv2018}) offers hope to democratize deep learning models to 
low powered devices. More efforts are needed to make deep learning models usable for IoT devices. In addition, the deployment of an already trained 
model in IoT devices can be challenging due to the wide varieties of IoT devices. IoT vendors can take this as an opportunity to crate container-based 
tools that do not have to rely on specific device configurations. 
\bf{IoT Developers} can benefit from such techniques and tools to make their IoT-based devices smart and secure. IoT 
developers can also use our results to learn about the state-of-the-art in security and ML-adoption in IoT ecosystem. 
The \bf{IoT Educators} 
can develop tutorials and documentation to teach security and ML basics IoT practitioners. The development and contents 
of such tutorials can be guided by the insights of developers' problems discussed in SO. This is important because as we noted a large number of ML-related 
discussions in SO are simply basic tutorials like definitions of supervised and unsupervised ML, etc. 
The \bf{IoT Researchers} can analyze the security and ML-adoption discussions to know about the specific challenges that IoT developers 
are facing based on their real-world experience. Such insights can be useful for the researchers to invent new techniques and tools, e.g., 
low-cost and less resource intensive ML models for IoT devices.  

\nd\bf{\ul{Threats to Validity.}}  \it{Internal validity} threats relate to our bias 
while conducting the analysis. We mitigated the bias by using automated techniques to pick security and ML-related discussions. 
The security discussions are picked using a deep machine learning model that shows an F1-score of 0.92. 
The ML-related discussions are picked based on a set of keywords that denote the different types of ML algorithms. 
\it{Construct validity} threats relate to the difficulty in finding
data to create our IoT security and ML-related sentences. While our data collection is based on automated approaches, we can miss 
relevant discussions that are not detected by our ML model to detect security or the discussions do not have any ML-specific keywords. 
However, our security detector model has high precision (0.92) and recall (0.93). Our ML-specific keywords also cover the different types of ML algorithms.   
\it{External validity} threats relate to the
generalizability of our findings. Our findings are based on SO. 
A detailed analysis of security and ML-adoption based on developer discussions in other online forum is our future work. 
\section{Related Work} \label{sec:background}
Related work can broadly be divided into \bf{Studies} to understand and \bf{Techniques} to create smart and secure IoT tools.

\nd\bf{\ul{Studies.}} Literature in IoT so far has focused on
surveys of IoT techniques and
architectures~\cite{Sethi-IoTArchitecture-JECE2017,Atzori-SurveyIoT-ComputerNetwork2010,Yang-StudyIoTArchitecture-ICMT2011,Fuqaha-IoTSurveyTechnologiesApplications-IEEECST2015},
the underlying middleware solutions (e.g.,
Hub)~\cite{Chaqfeh-ChallengesMiddlewareIoT-2012}, the use of big data analytics
to make smarter devices~\cite{Marjani-IoTDataAnalytics-IEEEAccess2017}, the
design of secure protocols and
techniques~\cite{Fuqaha-IoTSurveyTechnologiesApplications-IEEECST2015,Khan-IoTSecurityReview-FGCS2018,Frustaci-IoTSecurityEvaluation-IEEEIoTJournal2017,Zhang-IoTSecurityChallenge-SOCA2014}
and their applications on diverse domains (e.g.,
eHealth~\cite{Minoli-IoTSecurityForEHealth-CHASE2017}), the Industrial adoption
of
IoT~\cite{Liao-IndustrialIoT-IEEEIoT2018,Wang-TowardSmartFactoryIndustry4-ComputerNetwork2016},
and the evolution and visions related to IoT
technologies~\cite{Pretz-TheNextEvolutionInternet-IEEEMagazie2013,Sharma-HistoryIoT-ElsevierIoT2019,Chase-EvolutionIoT-TexasInstruments2013}.
The unauthorized inference of sensitive information from/among IoT devices 
is a prevalent concern\cite{Celik-IoTSensitiveInformationTracking-USENIX2018, Ding-IoTSafetyPhysicalInterfaction-CCS2018, Ho-IoTSmartLocks-ASIACCS2016}.
We are aware of no previous research that focused on understanding
IoT security and ML discussions in SO.  

\nd\bf{\ul{Techniques.}} IoT devices 
can be easy target for cyber threats~\cite{Zhang-IoTSecurityChallenge-SOCA2014,Frustaci-IoTSecurityEvaluation-IEEEIoTJournal2017,Wang-FearLoggingIoT-NDSS2018}. As such, significant research efforts are underway to improve IoT security. Automated IoT security and safety measures are
studied in Soteria \cite{Celik-IoTSafetySecurityAnalysis-USENIX2018},
IoTGuard~\cite{Celik-IoTDynamicEnforcementOfSecurity-NDSS2019}. Encryption and
hashing technologies make communication more secure and certified~\cite{Tedeschi-LikeSecureIoTCommunications-IEEEIoT2020}. Many authorization
techniques for IoT are proposed like SmartAuth~\cite{YuanTian-APIBot-ASE2017}. For smart home security, IoT security techniques are proposed like Piano~\cite{Gong-IoTPIANO-ICDCS2017},  smart authentication~\cite{He-RethinkIoTAccessControl-USENIX2018}, and cross-App Interference threat mitiation~\cite{Chi-SmartHomeCrossAppInference-DSN2020}. Session management and token verification are used in web security to
prevent intruder getting information. Attacks on Zigbee, an IEEE specification used to support interoperability can make IoT devices vulnerable~\cite{Ronen-IoTNuclearZigbeeChainReaction-SP2017,Wang-IoTAttackLandscapeZigbee-WiSec2020}. Smart gateway for IoT is proposed to tackle
malicious attack~ \cite{Hussain-IoTSecurityLink-DCOSS2019}. The adoption of machine learning algorithms for IoT devices is an emerging research area~\cite{Yousefi-EnergyEfficientClusteringIoT-CEE2020,Yousefi-IoTEfficientRoutePlanning-AHN2020,Yazdinejad-DroneIoTBlockchain-IEEEIoTJ2020}. 
To the best of our knowledge, ours is the first study 
that used automated techniques to analyze IoT security and ML discussions in SO. 

\section{Conclusions} \label{sec:conclusion}
The rapid adoption of IoT-based solutions 
has necessitated the needs to develop proper security and machine learning (ML) techniques for IoT devices and communications. 
As such, it is important to understand the
problems IoT developers discuss about their usage of security and ML tools and techniques in online technical forums
like Stack Overflow (SO), which is one of the most popular online forums for software developers. 
We studied the security and ML-related discussions at the sentence level in a dataset of 53K IoT posts from SO. 
We find that IoT developers discuss problems related to security and ML adoption, with concerns and discussions about 
security are much more prevalent than ML-specific adoption. We find that security discussions can be multifaceted that 
can range from securing data access/storage/transmission across devices and users as well as the incompatibilities of IoT devices 
with regards to the adoption of a security techniques/tools across platforms/devices. We also find that the resource constraints in 
the IoT devices make it challenging for IoT developers to adopt the recent advances in ML like deep learning models into the IoT devices.
Out findings can offer insights to various IoT stakeholders like IoT vendors and researchers to improve the state-of-the-art practices of 
security and ML-specific adoptions across the diverse IoT ecosystems.

\begin{small}
\bibliographystyle{abbrv}
\bibliography{consolidated}
\end{small}

\end{document}